\documentclass{iau}
\usepackage{graphicx}

\title[IAUS~304.~~Close pairs of galaxies with different activity levels] %% give here short title %%
{Close pairs of galaxies \\with different activity levels}

\author[T.~A.~Nazaryan et al.]   %% give here short author list %%
{T.~A.~Nazaryan$^1$,
A.~R.~Petrosian$^1$,
A.~A.~Hakobyan$^1$,\\
B.~J.~McLean$^2$,
 \and D.~Kunth$^3$}

\affiliation{$^1$Byurakan Astrophysical Observatory, 0213 Byurakan, Aragatsotn Province, Armenia \\
email: {\tt nazaryan@bao.sci.am} \\[\affilskip]
$^2$Space Telescope Science Institute, 3700 San Martin Drive, Baltimore, MD 21218, USA\\[\affilskip]
$^3$Institut d'Astrophysique de Paris, 98 bis Bd Arago, 75014 Paris, France
}

\pubyear{2013}
\volume{304}  %% insert here IAU Symposium No.
\pagerange{000--000}
\jname{Multiwavelength AGN Surveys and Studies}
\editors{A.~M.~Mickaelian \& D.~B.~Sanders, eds.}
\begin{document}

\maketitle

\begin{abstract}
We selected and studied 180 pairs with ${\rm d} V < 800~ km~ s^{-1}$
and $D\rm{p} < 60~ kpc$ containing Markarian (MRK) galaxies
to investigate the dependence of galaxies’ integral parameters,
star-formation (SF) and active galactic nuclei (AGN) properties on
kinematics of pairs, their structure and large-scale environments.
Projected radial separation $D\rm{p}$ and perturbation level $P$ are
better measures of interaction strength than ${\rm d}V$. The latter correlates with
the density of large-scale environment and with the morphologies of galaxies. Both galaxies in a
pair are of the same nature, the only difference is that MRK galaxies are usually brighter than
their neighbors. Specific star formation rates (SSFR) of galaxies in pairs
with smaller $D\rm{p}$ or ${\rm d } V$ is in average 0.5~dex higher than that of galaxies in pairs with larger $D\rm{p}$ or
${\rm d } V$. Closeness of a neighbor with the same and later morphological type increases the SSFR,
while earlier-type neighbors do not increase SSFR. Major interactions/mergers trigger SF and
AGN more effectively than minor ones. The fraction of AGNs is higher in more perturbed pairs
and pairs with smaller $D\rm{p}$. AGNs typically are in stronger interacting systems than star-forming
and passive galaxies. There are correlations of both SSFRs and spectral properties of nuclei
between pair members.
\end{abstract}

\firstsection % if your document starts with a section,
\section{Introduction}
\label{intro}

Interactions/mergers of galaxies are considered as important
processes influencing morphological, stellar and chemical evolution of galaxies.
Interactions/mergers can enhance star formation rates (SFRs) in galaxies
as is revealed by numerous observations of close pairs (e.g.
\cite[Larson \& Tinsley 1978]{larson78},
\cite[Lambas et al. 2003]{lambas03},
\cite[Li et al. 2008]{li08},
\cite[Ellison et al. 2008]{ellison08},
\cite[2013]{ellison13},
\cite[Hwang et al. 2010]{hwang10}
).
The main physical processes responsible are
gas inflow toward nuclear regions of galaxies due to global torques and,
probably, gas fragmentation into massive and dense clouds and rapid SF therein
(e.g.
\cite[Mihos \& Hernquist 1996]{mihos96},
\cite[Hopkins et al. 2013]{hopkins13}).
The triggering mechanism of AGN is often considered to be the same as that
of the enhanced nuclear SF
(e.g.
\cite[Ellison et al. 2008]{ellison08},
\cite[2011]{ellison11a},
\cite[Wild, Heckman \& Charlot 2010]{wild10},
\cite[Liu et al. 2012]{liu12}).

Large-scale environments, morphologies of both galaxies and mass ratios can affect on
frequency and efficiency of enhanced SF and AGN triggering by interactions and merging
(e.g.
\cite[Sol Alonso et al. 2006]{alonso06},
\cite[Hwang et al. 2010]{hwang10},
\cite[Scudder et al. 2012]{scudder12a},
\cite[Li et al. 2008]{li08},
\cite[Ellison et al. 2008]{ellison08},
\cite[2010]{ellison10},
\cite[2011]{ellison11a},
\cite[Cox et al. 2008]{cox08},
\cite[Lambas et al. 2012]{lambas12}).
The facts, that not all galaxies with high SFR are interacting ones,
as well as that not all interacting galaxies have high SFR,
support the hypothesis that internal properties of galaxies are also
an important factor determining enhanced SF
(\cite[Lambas et al. 2003]{lambas03},
\cite[Sol Alonso et al. 2006]{alonso06}).

The aim of this study is to investigate the connections
between interaction with a close neighbor
and nuclear activity and/or enhanced SF in galaxy pairs
using the sample of close neighbors of MRK galaxies from \cite[Nazaryan et al. (2012)]{nazaryan12}.
The complete study is presented in  \cite{nazaryan14a}.
A similar study of Second Byurakan Survey galaxies is presented in \cite{nazaryan14b}.

\section{Sample}
\label{sample}

The starting point to create our sample of pairs is the catalog of MRK galaxies.
The original catalog (\cite[Markarian et al. 1989]{markarian89}) features 1545 bright galaxies
mostly having starburst properties and/or AGNs.
In \cite{petrosian07}, homogeneously measured parameters of MRK galaxies are presented.
Results of a close neighbors search for MRK galaxies
are published in \cite[Nazaryan et al. (2012)]{nazaryan12}.
For the current study, the subsample of 217 pairs of galaxies containing at least one MRK galaxy is selected.
For this study, we classified morphologies of MRK galaxies and neighbors
by using SDSS color images. We also classified sample pairs in terms of morphological perturbations
(see e.g.
\cite[Liu et al. 2012]{liu12},
\cite[Lambas et al. 2012]{lambas12})
by 4 levels: $P = 0$: unperturbed pairs, $P = 1$: slightly perturbed, $P = 2$: highly perturbed, $P = 3$: mergers.
We described the large-scale environments of each pair by average large-scale density ${\it \Sigma}$.
For statistics we also included SFRs of galaxies and nuclear emission-line classification (BPT) of their SDSS spectra.

\section{Statistics and discussions}
\label{statistics}

The dependence of visually detected perturbation level $P$ on ${\rm d}V$ and $D\rm{p}$ is worthy to mention.
Disturbance level $P$ correlates only with $D\rm{p}$,
while ${\rm d}V$ and $D\rm{p}$ do not correlate with each other.
The closer pairs are more disturbed.
This is the result of different nature of $D\rm{p}$ and ${\rm d}V$.
Pairs with larger ${\rm d}V$ correspond to environments with higher densities ${\it \Sigma}$.
Therefore while $D\rm{p}$ is a measure of interaction strength,
the variation of ${\rm d}V$ mainly reflects change of large-scale environments.
The SSFRs of barred galaxies do not differ from those of unbarred ones significantly.

We compared properties of neighbors with those of MRK galaxies.
Mean absolute \emph{r}~mag of neighbors is fainter by 0.9 mag than that of MRK galaxies.
Morphologies of neighbors correlate significantly with those of MRK galaxies and are slightly later.
Neighbor galaxies have the same distributions by SSFRs, BPT types and colors, and the same fraction of barred galaxies as MRK galaxies.
This is a result of a correlation between properties of galaxies in pairs.
However, because of magnitude limitation of Markarian survey,
MRK galaxies are usually the brightest members of pairs.

\begin{figure}[t]
\centering
\includegraphics[width=0.48\hsize]{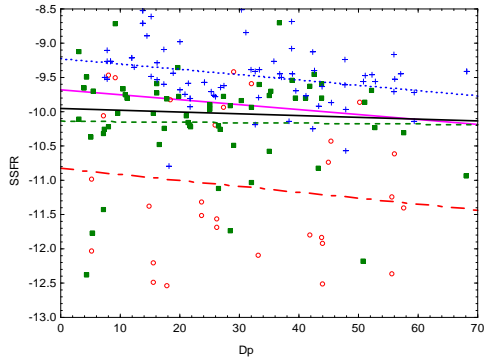}\\
\parbox{0.48\hsize}{\caption{SSFR vs. $D\rm{p}$ for subsamples
of early-types (red blank circles, dashed-dotted
line), early spirals (green filled squares,
dashed line), and late spirals and
irregulars (blue crosses, dotted line). Two
best-fit lines for all galaxies (black bottom solid
line) and AGN--removed sample (purple upper
solid line) are drawn.}\label{SSFRvsDpmorph}}
\end{figure}
\begin{figure}[t]
\begin{minipage}[t]{0.48\linewidth}
\centering
\includegraphics[width=1\hsize]{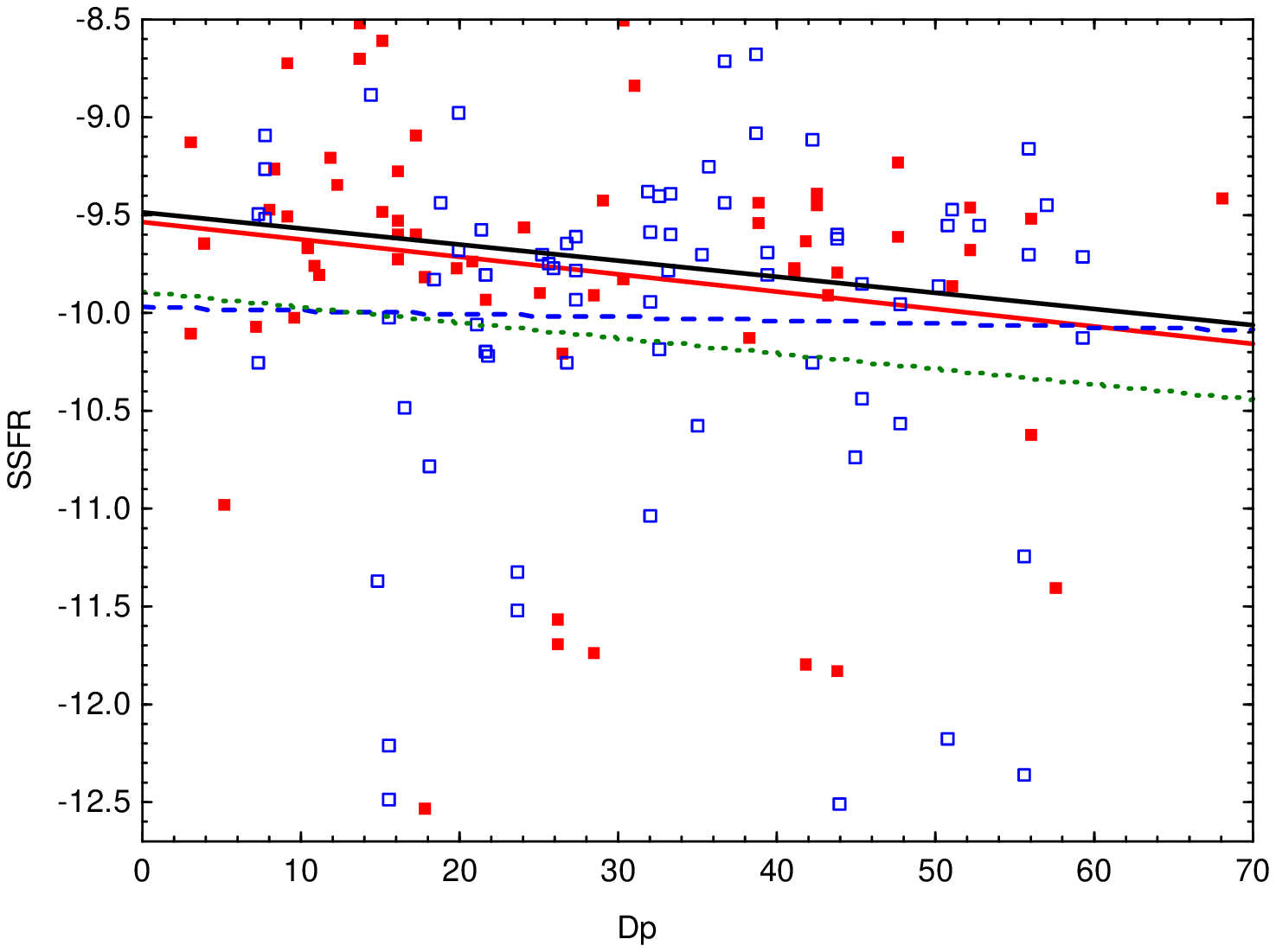}
\caption{SSFR vs. $D\rm{p}$ for major
interactions (red filled squares, solid line
best fit), minor interactions (blue blank
squares, dashed line), brightest
components of major interactions
(black upper solid line), and brightest
components of minor interactions
(green dotted line).}
\label{SSFRvsDpmajmin}
\end{minipage}
\hspace{0.04\linewidth}
\begin{minipage}[t]{0.48\linewidth}
\centering
\includegraphics[width=1\hsize]{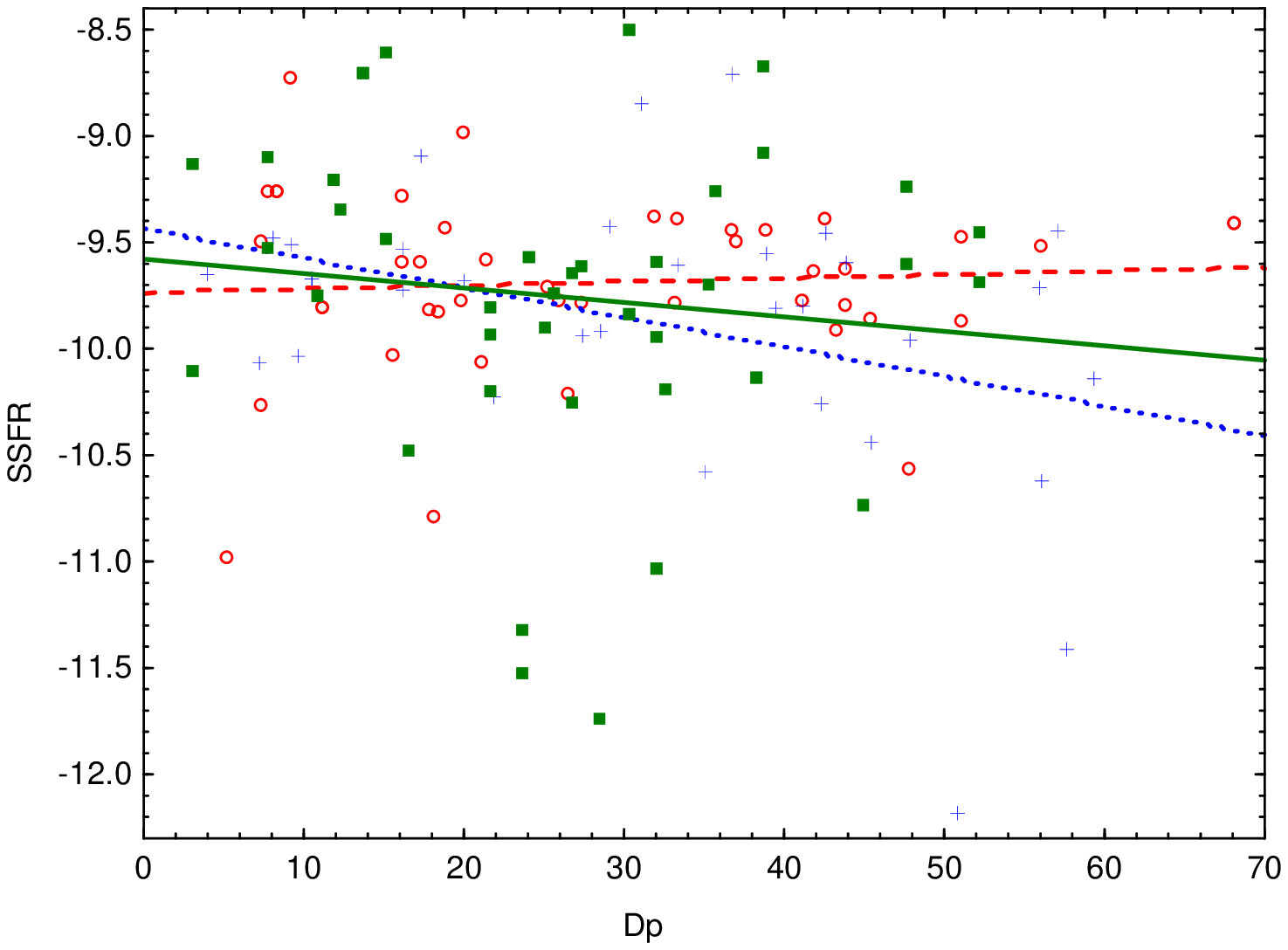}
\caption{SSFR vs. $D\rm{p}$ for galaxies with
relatively earlier type neighbors (red blank
circles, dashed line), same type
neighbors (green filled squares, solid line),
and relatively later type neighbors
(blue crosses, dotted line).}
\label{SSFRvsnmorph}
\end{minipage}
\end{figure}

The main parameters describing interactions are ${\rm d}V$, $D\rm{p}$, and $P$.
Without considering morphologies,
there is a 0.7~dex increase of SSFRs from larger ${\rm d}V$ to smaller ones.
However the variance of SSFRs because of morphologies is much larger (more than 2.5~dex).
In \cite{ellison10}, it was shown that
pairs in denser environments have larger ${\rm d}V$.
Because of morphology--density relation, ${\rm d}V$ is also biased by morphologies:
early-type galaxies have about 3 times more ${\rm d}V$ than irregulars,
so most of the SSFR vs. ${\rm d}V$ dependence is because of morphology--SSFR dependence
and does not reflect pure interaction.
Therefore, it is essential to take into account morphologies of galaxies
when discussing their SSFRs and interactions.
Grouping by morphologies weakens SSFR vs. ${\rm d}V$ relation,
but there still remains some variance, maximal for early spirals ($0.4 - 0.5$~dex).
This result is in agreement with modeling, e.g. \cite{dimatteo08}
showing that strong starbursts during interactions are rare
and that typical enhancement of SFR is less than 5 times.
Figure~\ref{SSFRvsDpmorph} shows the dependence of SSFR on $D\rm{p}$ of a pair in AGN--removed sample.
In our sample $D\rm{p}$ is not biased by morphologies compared to ${\rm d}V$.
After grouping by morphologies we have about 0.5~dex increase of SSFRs
in closer pairs.

We studied the impact of the luminosity ratio of pair members on SSFR increase
by considering major
($log(L_{\rm{bright}}/ L_{\rm{faint}})\leq 0.6$)
and minor (the rest) pairs.
The major interactions are more effective in triggering SF than minor ones (see Fig.~\ref{SSFRvsDpmajmin}),
there is a 0.5~dex increase of SF in major interactions,
while there is no trend among minor interactions.
This results are in agreement with both observational and modeling data
(\cite[Lambas et al. 2003]{lambas03},
\cite[2012]{lambas12},
\cite[Cox et al. 2008]{cox08})
suggesting that the merger mass ratio is
an important parameter defining the effectiveness of the tidal forces.

The impact of morphology of neighbor galaxy on SSFR is shown in Fig.~\ref{SSFRvsnmorph}.
Earlier-type neighbors do not increase of SSFR,
while the same- and later-type neighbors increase SSFR.
The extra SSFR is maximal if the neighbor is of later morphological type (0.8~dex).
Previous papers, e.g. \cite{hwang10}, also obtained similar result,
supporting the scenario where neighbor of later type
can not only trigger gas inflow in earlier type galaxy,
but also be an additional source of gas fuel.

The fraction of AGN galaxies in less separated pairs is larger than that in more separated pairs.
Fraction of AGNs increases by about 7 times when going from unperturbed pairs ($P=0$)
to mergers ($P=3$).
This result indicates that
while both AGN and SF can be triggered by interactions,
AGNs ``prefer'' stronger interactions.
Different timings of starbursts and AGN events can explain this result
(\cite[Ellison et al. 2008]{ellison08},
\cite[Wild et al. 2010]{wild10},
\cite[Hopkins 2012]{hopkins12}).
The fraction of AGNs in major interactions is about 4 times larger than
that in minor ones.

We studied BPT--BPT correlations between pair members.
There is a tendency to have an increased probability of a neighbor with the same BPT type.
Especially that is noticeable regarding passive galaxies and AGNs:
passive galaxies are more likely to be found near another passive galaxy,
AGNs are more probable to be near another AGN.
We explain these correlations between BPT types as a result of two main factors:
first, the correlation between morphologies of galaxies, and, second, the increased likelihood of pair members
to have nuclear activity of same types when interacting strongly.

\begin{acknowledgement}
A.~R.~P. and A.~A.~H. acknowledge the hospitality of the Institut d'Astrophysique de Paris
(France) during their stay as visiting scientists supported by the Collaborative Bilateral Research
Project of the State Committee of Science (SCS) of the Republic of Armenia and the French
Centre National de la Recherch\'{e} Scientifique (CNRS).
This work was made possible in part by
a research grant from the Armenian National Science and Education Fund (ANSEF) based in
New York, USA.
\end{acknowledgement}


\begin{thebibliography}{}


\bibitem[Cox et al. (2008)]{cox08}
{Cox,~T.~J., Jonsson,~P.,~Somerville,~R.~S., et al.} 2008, \textit{MNRAS}, 384, 386

\bibitem[Di Matteo et al. (2008)]{dimatteo08}
{Di~Matteo,~P., Bournaud,~F., Martig,~M., et al.} 2008, \textit{A\&A}, 492, 31

\bibitem[Ellison et al. (2008)]{ellison08}
{Ellison,~S.~L., Patton,~D.~R., Simard,~L.,~ \& ~McConnachie,~A.~W.} 2008, \textit{AJ}, 135, 1877

\bibitem[Ellison et al. (2010)]{ellison10}
{Ellison,~S.~L., Patton,~D.~R., Simard,~L., et al.} 2010, \textit{MNRAS}, 407, 1514

\bibitem[Ellison et al. (2011)]{ellison11a}
{Ellison,~S.~L., Patton,~D.~R., Mendel,~J.~T., \& Scudder,~J.~M.} 2011, \textit{MNRAS}, 418, 2043

\bibitem[Ellison et al. (2013)]{ellison13}
{Ellison,~S.~L., Mendel,~J.~T., Scudder,~J.~M., et al.} 2013, \textit{MNRAS}, 430, 3128

\bibitem[Hopkins (2012)]{hopkins12}
{Hopkins,~P.~F.} 2012, \textit{MNRAS}, 420, L8

\bibitem[Hopkins et al. (2013)]{hopkins13}
{Hopkins,~P.~F., Cox,~T.~J., Hernquist,~L., et al.} 2013, \textit{MNRAS}, 430, 1901

\bibitem[Hwang et al. (2010)]{hwang10}
{Hwang,~H.~S., Elbaz,~D., Lee,~J.~C., et al.} 2010, \textit{A\&A}, 522, A33

\bibitem[Lambas et al. (2003)]{lambas03}
{Lambas,~D.~G., Tissera,~P.~B., Alonso,~M.~S., \& Coldwell,~G.} 2003, \textit{MNRAS}, 346, 1189

\bibitem[Lambas et al. (2012)]{lambas12}
{Lambas,~D.~G., Alonso,~S., Mesa,~V., \& O'Mill,~A.~L.} 2012, \textit{A\&A}, 539, A45

\bibitem[Larson \& Tinsley (1978)]{larson78}
{Larson,~R.~B., \& Tinsley,~B.~M.} 1978, \textit{ApJ}, 219, 46

\bibitem[Li et al. (2008)]{li08}
{Li,~C., Kauffmann,~G., Heckman~T.~M., et al.} 2008, \textit{MNRAS}, 385, 1903

\bibitem[Liu, Shen \& Strauss (2012)]{liu12}
{Liu,~X., Shen,~Y., \& Strauss,~M.~A.} 2012, \textit{ApJ}, 745, 94

\bibitem[Markarian et al. (1989)]{markarian89}
{Markarian,~B.~E., Lipovetsky,~V.~A., Stepanian,~J.~A., et al.} 1989, \textit{Comm. SAO}, 62, 5

\bibitem[Mihos \& Hernquist (1996)]{mihos96}
{Mihos,~J.~C., \& Hernquist,~L.} 1996, \textit{ApJ}, 464, 641

\bibitem[Nazaryan, Petrosian \& McLean (2012)]{nazaryan12}
{Nazaryan,~T.~A., Petrosian,~A.~R., \& McLean,~B.~J.} 2012, \textit{Astrophys.}, 55, 448

\bibitem[Nazaryan et al. (2014)]{nazaryan14a}
{Nazaryan,~T.~A., Petrosian,~A.~R., Hakobyan,~A.~A.,  et al.} 2014, \textit{Astrophys.}, 57, 14

\bibitem[Nazaryan (2014)]{nazaryan14b}
{Nazaryan,~T.~A.} 2014, \textit{Astrophys.}, 57, 50

\bibitem[Petrosian et al. (2007)]{petrosian07}
{Petrosian,~A., McLean,~B., Allen,~R.~J., \& MacKenty,~J.~W.} 2007, \textit{ApJS}, 170, 33

\bibitem[Scudder, Ellison \& Mendel (2012a)]{scudder12a}
{Scudder,~J.~M., Ellison,~S.~L., \& Mendel,~J.~T.} 2012, \textit{MNRAS}, 423, 2690

\bibitem[Sol Alonso et al. (2006)]{alonso06}
{Sol~Alonso,~M., Lambas,~D.~G., Tissera,~P., \& Coldwell,~G.} 2006, \textit{MNRAS}, 367, 1029

\bibitem[Wild, Heckman \& Charlot (2010)]{wild10}
{Wild,~V., Heckman,~T., \& Charlot,~S.} 2010, \textit{MNRAS}, 405, 933




\end{thebibliography}
\end{document}